\documentclass[sigconf]{acmart}
\usepackage{subfigure}
\usepackage{verbatim}
\AtBeginDocument{%
  \providecommand\BibTeX{{%
    \normalfont B\kern-0.5em{\scshape i\kern-0.25em b}\kern-0.8em\TeX}}}

\setcopyright{acmcopyright}
\copyrightyear{2018}
\acmYear{2018}
\acmDOI{10.1145/1122445.1122456}

\acmConference[FDG '20]{The Fifteenth International Conference on the Foundations of Digital Games}{September 15--18, 2020}{Bugibba, Malta}
\acmBooktitle{The Fifteenth International Conference on the Foundations of Digital Games (FDG '20), September 15--18, 2020, Bugibba, Malta}
\acmPrice{15.00}
\acmISBN{978-1-4503-XXXX-X/18/06}

\begin{document}

\title{Data-Driven Game Development: Ethical Considerations}
%Leveraging Transparency Towards More Ethical Player Modeling}


\author{Magy Seif El-Nasr} 
\affiliation{%
  \institution{Northeastern University}
}
\email{magy@northeastern.edu}

\author{Erica Kleinman} 
\affiliation{%
  \institution{Northeastern University}
}
\email{kleinman.e@husky.neu.edu}

\begin{abstract}
In recent years, the games industry has made a major move towards data-driven development, using data analytics and player modeling to inform design decisions. Data-driven techniques are beneficial as they allow for the study of player behavior at scale, making them very applicable to modern digital game development. However, with this move towards data driven decision-making comes a number of ethical concerns. Previous work in player modeling \cite{mikkelsen2017ethical} as well as work in the fields of AI and machine learning \cite{bostrom2014ethics,o2016weapons} have demonstrated several ways in which algorithmic decision-making can be flawed due to data or algorithmic bias or lack of data from specific groups. Further, black box algorithms create a trust problem due to lack of interpretability and transparency of the results or models developed based on the data, requiring blind faith in the results. In this position paper, we discuss several factors affecting the use of game data in the development cycle. In addition to issues raised by previous work, we also raise issues with algorithms marginalizing certain player groups and flaws in the resulting models due to their inability to reason about situational factors affecting players' decisions. Further, we outline some work that seeks to address these problems and identify some open problems concerning ethics and game data science.
\end{abstract}




\maketitle

%%%%%%%%%%%%%%%%%%%%%%%%%%%%%%%%%
%%%%%%% PAPER BEGINS HERE %%%%%%%
%%%%%%%%%%%%%%%%%%%%%%%%%%%%%%%%%

\section{Introduction}
Data Science is becoming a tool-set used in many applications, including recruitment, marketing, design, workflow analysis, and policy making --- affecting people as consumers as well as producers of technology. Within game development, specifically, game data science and analytics have been used to tune and adapt design \cite{el2016game,hauge2014implications,harrison2012creating}, assess workflow \cite{zoeller2013game}, and recommend teams or items to players \cite {chen2018q,chen2018art}. With such an increase in the use of data within many of the processes affecting users, organizations, and developers, several researchers questioned the ethical implications of such practices. In her book, \textit{Weapons of Math Destruction}, O'Neil discusses the issues stating: "consequently, racism is the most slovenly of predictive models. It is powered by haphazard data gathering and spurious correlations, reinforced by institutional inequities, and polluted by confirmation bias \cite{o2016weapons}, [P. 23]." 

O'Neil isn't the first to express concerns regarding the moral and ethical impact of Artificial Intelligence (AI) on our society. Bostrom and Yudkowsky \cite{bostrom2014ethics} give a detailed overview of the types of ethical considerations that need to be taken into account as algorithms are increasingly charged with tasks that have legitimate social impact. They warn that lack of data, or incorrect selection of data, can inject bias into the output. Further, because the inner workings of many Machine Learning (ML) systems are often unknown to most, it can be difficult to know why an AI made a certain decision. This becomes increasingly problematic when those decisions have a negative social impact, such as denying mortgages to African American applicants \cite{bostrom2014ethics}, or misidentifying them as gorillas --- a bias in the google image identification algorithm reported in 2015 \cite{vincent2018google}.

These concerns also raise issues of users' trust or mistrust in the system. Studies show that the people who would, and do, interact with these algorithms in the wild rarely trust them \cite{eslami2019user,binns2018s}. Binns et al. \cite{binns2018s} conducted a study in which they found that many participants felt it was dehumanizing or impersonal to allow a machine to make decisions for humans. They also stated concerns over their inability to understand how the systems arrived at their answers, generalization, and statistical inference \cite{binns2018s}. Eslami et al.'s \cite{eslami2019user} work examining the Yelp algorithm found that many users did not trust the opaque system and expressed that they felt it was discouraging or even impeding their freedom of speech, with many users demanding transparency. Users' concerns regarding transparency and explainability are also echoed by Siau and Wang \cite{siau2018building} as critical factors affecting whether or not people can trust AI.

So what does that mean for games? As the game industry and academic research starts to streamline processes for the use of data in many aspects of game development, design, and research, they also will face similar issues and concerns regarding privacy, bias, unfairness, and marginalization of specific groups of players, which may raise issues of trust or distrust among the community. This may have more severe impact in applications of games in education, health, or others \cite{canossa2014reporting}. Mikkelsen et al. \cite{mikkelsen2017ethical} discussed numerous ethical drawbacks to player modeling. For example, they raised the issue that algorithms designed to increase monetization and prevent churn do not consider the financial status of the player, meaning they risk trapping the player in a predatory cycle, spending money they do not have. Automated banning systems can benefit the player base by punishing inappropriate behavior, but must be transparent in terms of what they target, so players can intentionally avoid them. Recommendation systems and adaptive systems may not properly accommodate all players equally, resulting in unpleasant gaming experiences for minorities or non-expert players. Recommendation systems may also potentially discriminate against players with less purchasing power. They also warn that using models during the balancing process can result in gameplay adjustments that do not apply to the entire player-base. In addition to all of these concerns, there is also the underlying issue of privacy \cite{mikkelsen2017ethical}. 

Like O'Niel \cite{o2016weapons}, Mikkelsen et al. \cite{mikkelsen2017ethical} recommend the use of transparent models to address some of these issues. 
While transparency is important, there needs to be a concrete methodological approach that includes transparency as part of the process when addressing these issues. A fair body of work has been dedicated to creating algorithms and algorithmic methodologies that are considered more ethical or fair \cite{corbett2018measure,stahl2018ethics,berendt2019ai}. Zhu et al. \cite{zhu2018value} and Roselli et al. \cite{roselli2019managing} attempted to resolve these ethical drawbacks by increasing human involvement in the modeling process in order to increase transparency, explainability, and auditability. Zhu et al. \cite{zhu2018value} developed a novel approach to algorithm design they call ``Value-Sensitive Algorithm Design". They sought to address two concerns 1) a lack of critical engagement with users and stakeholders, and 2) a reliance on historical judgements rather than insights into how the world can be improved. They incorporated stakeholder knowledge and feedback early in development in order to avoid bias or undermine stakeholder values in the algorithm's design \cite{zhu2018value}. They use an iterative design process in which the algorithm is tuned based on feedback to remove unintended biases. Further, Roselli et al. \cite{roselli2019managing} presented a series of steps that can be taken when training a model to reduce bias. They emphasized that substantial evidence should be present to back up a hypothesis and that training data should not be overly curated and guarded against manipulation, to avoid bias. They also pushed for a feedback loop in which the outcomes are checked against predictions and negative feedback loops are avoided by comparison with external evidence \cite{roselli2019managing}. While this work presents a great stride towards a methodology that address such ethical issues, such processes have not been discussed within the context of the game data science pipeline. 

These interpretability concerns have also given rise to the field of Explainable AI (XAI), which employs various methods and techniques to ensure that the outputs of an AI system are explainable and interpretable by humans \cite{doran2017does,miller2017explainable,holzinger2018machine,wang2019designing}. Work in XAI sought to normalize what ``explainable" means \cite{doran2017does} such that AI systems can be designed and implemented in a consistent manner. Further, work has sought to connect facets of ``explainability" with facets of human reasoning to better guide XAI development \cite{wang2019designing}. However, there are concerns that much of the explainability guidelines are more focused on explaining the systems to the developers rather than to users \cite{miller2017explainable}. Further, te work in this field is still in its beginning and more work needs to be done to make such approaches useful for the game data science pipeline. 

While there has been a strong push for more ethical and explainable AI, little work has been done in the field of games. To make matters more difficult, analyzing game data is an incredibly complicated process. Game data is often influenced by numerous factors, including, but not limited to, individual differences among players and the situational nature of gameplay. In terms of individual differences, there are various player differences that will naturally impact the way any given player approaches a game. These range from cultural differences \cite{sun2017motivation,parshakov2015success} to personality differences \cite{bunian2017modeling, chen2015modeling} to skill based differences \cite{drachen2014skill,chen2015modeling}. At the same time, decision-making in games is heavily informed and influenced by the context in which it occurs \cite{ahmad2019modeling}. This results in highly situational and individually varied data that cannot easily be aggregated or distributed for analysis without introducing some risk of bias or misinterpretation into the process.

In this position paper, we argue that issues impacting data-driven techniques and AI concerning fairness and ethics impact game data science as well. We expand on the concerns discussed by others echoing issues of individual differences and situational decision-making. We then build on previous works' discussions of ethical concerns in game data by enumerating some new methods being developed and explored in games that can begin to address these concerns. We then provide takeaways that we believe can help inform more ethical data driven game design. We conclude by discussing the issues that need to be addressed by game data scientists, the open problems and their implications on our field. We hope this position paper can be a start to the discussion around game data science and ethics --- as such discussion has already been underway in the fields of ML and AI. 

\section{Ethical Concerns with the use of Game Data}
Games, as is the case with other media, is a genre of entertainment that have long been concerned with ethics due to the potential for the medium to influence consumers \cite{versteeg2013ethics,sicart2010values,hodent2019cognitive,hodent2019cognitive}. There have been many discussions regarding ethics in game development, ranging from game content and exposing players to violence \cite{smith2003popular,mccormick2001wrong,waddington2007locating} to the role that games can play as vehicles to inspire reflection and thought \cite{sicart2009banality,sicart2010wicked,jorgensen2016positive,payne2014war}. We are particularly interested in the ethical use of data in the game development cycle, and thus we will focus the discussion here on this area. 

\subsection{Monetization and Its Impact on Players}
In 2017, EA's \emph{Star Wars Battlefront 2} \cite{starwars} announced a controversial lootbox system that had many claiming the game was ``pay to win" \cite{staff2017star}. The controversy directed attention to the monetization strategies of games and their tendency towards micro-transactions. Naturally, this gave rise to ethical concerns regarding the classification of such game elements as gambling \cite{drummond2018video}. Since then, a fair amount of work has examined the classification of lootboxes and the connection between them and gambling. A large scale study by Zendle and Cairns \cite{zendle2018video} found correlations between problematic gambling behaviors and the purchasing of lootboxes with real money. However, their study did not infer causation. In a more recent work Nielsen and Grabarczyk \cite{nielsen2019loot} developed a formal classification system for lootboxes (renamed ``random reward mechanisms") in which they break them down into categories, with only one type linked to gambling.

Much work still seeks to understand what motivates players to spend real money in otherwise free to play environments. Hamari et al. \cite{hamari2017players} performed a large scale study to address this issue. They identified several factors, including unobstructed play, social interaction, competition, and unlocking content. The use of monetization techniques have been criticized due to their use of player data to optimize on these factors. Mikkelsen et al. \cite{mikkelsen2017ethical}, for example, discussed how data can be used to determine the best way to encourage a player to make a purchase within a game. They then discussed that such data does not include any relevant information about the player, such as whether or not they are financially able to make such a purchase. The result is a predatory cycle, fueled by a desire for profit. Further, they illustrated how a player who makes financially responsible decisions, by not spending what they do not have, may be discriminated against by recommendation systems \cite{mikkelsen2017ethical}. 

These examples make it clear that the ethical concerns surrounding game monetization are closely linked to the ethical concerns surrounding game data, as it is such data that is often used to design free-to-play monetization systems. In her GDC talk, Celia Hodent warned that there is a lack of understanding of how design decisions may be impacting users' purchasing habits. These ``dark patterns", as she referred to them, include guilt tripping players into paying to make a character happy, or making it difficult to play for free \cite{hodent2019cognitive}. 

Despite these concerns, a study by Alha et al. \cite{alha2014free} found that most designers were only concerned about transactions being problematic in certain contexts, e.g., in the vicinity of children. They felt that free-to-play games with micro-transactions were no less ethical than a game you paid for upfront. However, issues of bias and recommendation systems' marginalization of particular players may be hidden in the black-box methods used.  

\subsection{Retention = Addiction}
As the games industry focuses more and more on games that rely on subscriptions, repeated or long term play, and in-game purchases, research on the use of data began to focus on how to keep people in the game longer \cite{bergstrom2019moving}. Various AI and player modeling techniques are used with the intent of achieving this goal. Xue et al. \cite{xue2017dynamic} developed a probabilistic model for dynamic difficulty adjustment to increase player engagement across levels. Hadiji et al. \cite{hadiji2014predicting} developed a set of classifiers that can be used to predict churn in ``freemium" (free to play, pay for bonuses) games. Additionally, Peterson et al. \cite{petersen2017evaluating} used a mixed methods approach to evaluate the on-boarding phase of free-to-play mobile games as they relate to the churn rate of new players. These are only a few of many works that sought to maximize player retention. However, there is a fine line between keeping players engaged in a game and stimulating addiction.

Hodent addressed addiction in her ethics talk, illustrating that, while concerns were valid, and there are those who need help, there was little evidence that video game 'clinical' addiction actually occured \cite{hodent2019cognitive}. However, Hodent drew on the clinical definition of addiction in her discussion, and the exact term used to diagnose patients in a clinical setting may not apply to most cases discussed for video games. There were nevertheless reported harmful behavioral signs among the game playing population, such as neglecting school, work, or family responsibilities \cite{wan2006psychological,mozelius2016gaming}. For example, a Korean couple allowed their child to starve to death while they played an online game \cite{salmon2010jail}. Additionally, studies conducted by Gr{\"u}sser et al. \cite{grusser2006excessive} showed that addictive gaming habits do not exist solely in correlation with monetization in games, and warned designers to be aware of how they entice players to continue playing, to avoid promoting damaging behavior. This is a particularly difficult task, as keeping players engaged is the overall goal of game design, and data is a helpful aid in achieving such a goal at scale. More work and discussion is needed on this dimension. On the flip side, there is also much data on players' game use and behavior which can be used to moderate play sessions. 

\subsection{Representation in Data}
As Mikkelsen et al. \cite{mikkelsen2017ethical} pointed out, it is difficult to create models that correctly account for everyone, and this becomes even more true when dealing with games that have large international communities and a broad user base. Previous work \cite{sun2017motivation,parshakov2015success} demonstrated that different location-based contexts will result in different gameplay motivations and behavior. This then raises a valid concern that aggregated data will not capture these differences. Further, data-sets that are drawn from a single region or skill level will result in a lack of representation in the data, and a model that is biased towards a particular group. This is an issue we will expand upon later in the paper. 

\subsection{Player Modeling}
Despite the fact that player modeling has grown in popularity \cite{smith2011inclusive,geisler2004integrated,hooshyar2018data}, the work on the ethical issues surrounding game AI and player modeling is sparse at best. Yannankis et al. \cite{yannakakis2013player} gave a detailed overview of the benefits and applications of player modeling. In discussing further work, they illustrated concerns that echo those regarding AI ethics, such as privacy, stereotypes, and censorship. Further, concerns regarding player's individual differences, and how models can adapt to these differences, are prominent. Hooshyar et al. \cite{hooshyar2019systematic} cite individual prediction as one of the challenges for player modeling in educational games, among others. Charles and Black \cite{charles2004dynamic} developed a framework for dynamic player modeling that attempted to address this by allowing the model to reclassify players as their performance drifted. 

However, individual differences are only one of several ethical issues within the field of player modeling. Mikkelsen et al. \cite{mikkelsen2017ethical} build on the work of ethics in AI to clearly articulate the ethical and moral shortcomings of player modeling as it relates to a variety of applications within games, including churn, recommendation systems, balancing, and matchmaking. In this paper, we echo some of these concerns and expand on them towards the development of methods that can account for these issues and biases. 

\section{Other Ethical Issues with Data-Driven Game Development}
There are other concerns regarding data-driven game development that have not been discussed extensively by previous work, but have been apparent in studies regarding player behavior in games. In this section we will discuss two issues we believe are important to the discussion here, particularly: Individual Differences and Situational Factors.

\subsection{Individual Differences = Error or Outliers?}
O'Neil \cite{o2016weapons} describes Simpson's Paradox --- a paradox that holds true with much of the work on game data. In her book, she states: ``Simpson’s Paradox: when a whole body of data displays one trend, yet when broken into subgroups, the opposite trend comes into view for each of those subgroups \cite{o2016weapons}[p. 136]." 

Games are dynamic and complex environments that rarely force players to solve problems in a single way. And previous work has found that partitioning players differently can bring about different results \cite{canossa2018like}. There are usually numerous influences that impact the way one approaches gameplay ranging from culture \cite{bialas2014cultural} to gender \cite{martey2014strategic,shen2016men} to expertise in game types or gameplay \cite{drachen2014skill,chen2015modeling,erfani2010effect}. The result is a myriad of vastly different play behaviors that players exhibit when playing a game --- such variation results in data skews. For example, expert players will play more levels and play longer, minimizing the data collected from non-expert players or players who give up early \cite{chen2015modeling}. Further, there is often a dominant strategy that will be employed by many players --- especially players who are used to specific types of games or puzzles. Due to ease, efficiency, or popularity, these dominant strategies are likely to overwhelm player data, and overshadow any uncommon (or creative) strategies that may exist at the fringes of the set. Such skews and lack of data from particular groups is challenging for automated modeling techniques, such as predictive classification or clustering algorithms. 

To illustrate the existence of such individual differences, Nguyen et al. \cite{nguyen2014strategy} devised an inverse reinforcement learning technique to predict policies, i.e. learned problem solving patterns, from game data logs. They used a game called \emph{Wuzzit Trouble} developed and published by BrainQuake Games \cite{wuzzit}. In their visualization of players' patterns, they showed how players varied in what rewards they were trying to maximize while playing the game. Further, they also showed that players varied in their goals and have, in fact, shifted strategies as they were playing due to lack of or changing opportunities and their own learning process. They concluded that inverse reinforcement learning techniques were not yet mature enough to address such variations in the data.   

Further, in our own work, we explored variation in problem solving behaviors through data collected from an educational game \cite{jemmali2020maads}. In our analysis of problem solving behaviors, we visualized all patterns to see how they varied over time. The visualization shows similar results to Nguyen et al.'s work \cite{nguyen2014strategy}. Similar to the previous example, the resulting visualizations indicated that there are several most-popular trajectories to solve the levels. However, there were also many diverse trajectories that existed in the data and were visualized, many of which belong to players who struggled with the debugging process. This is where the significance of individual differences really comes into play. Because this work deals with player performance in an educational game, and because the trajectories of those who were successful are the most common, aggregation or using vanilla ML techniques could marginalize players who need the most design adjustments or help. 

Such skews in the data become problematic when dealing with massive multiplayer online games, where the common strategy is often colloquially referred to as the ``meta". The ``meta" is followed most closely by the most dedicated players, who will also make up the majority of the data, simply due to their lengthy play time. Casual players, who are more likely to engage ``non-meta" strategies are thus more likely to be drowned out in the data, and overlooked by analysis. This can result in design changes that confuse or frustrate a large portion of a game's player-base, or, in some cases, provide unintentional advantages to those engaging deviant strategies that went unnoticed due to aggregation. 

These examples clearly illustrate how varied player behavior can be, even in simple games. When gameplay data is analyzed by traditional machine learning or statistical techniques, uncommon strategies or behaviors are frequently overlooked. However, if grouped by personality or other grouping factors, some of these patterns can be picked up by a modeling algorithm, e.g., \cite{bunian2017modeling}. Either to help those who need it, maintain balance, or improve quality of life, it is critical that game data analysis is able to capture individual differences. However, the aggregation, feature extraction, or generation methods used by most state-of-the-art techniques result in marginalizing individual with different play patterns. 

\subsection{What about Situational Factors ?}
In addition to overshadowing individual differences, most analytics or data science techniques rarely preserve the context in which actions or decisions occurred. However, player behavior is made up of situated actions \cite{suchman1987plans}, and it is difficult if not impossible to truly comprehend the strategic behavior of players without knowing the context that drove them to action.

Previous work by Ahmad et al. \cite{ahmad2019modeling} demonstrated the importance of spatio-temporal context to understanding player strategies in \emph{Defense of the Ancients 2} (DotA2) \cite{dota}. They used a human-in-the-loop methodology to apply behavioral labels to log data displayed in a spatio-temporal visualization system. This method allowed them to see when and where player actions occurred, which facilitated the capture of the state of the game environment when a behavior was taken \cite{ahmad2019modeling}. As a result, they were able to generate a more context-aware model of player behavior that allowed them to identify how strategies of winning and losing teams varied as a game progressed \cite{ahmad2019modeling}.

DotA2 players engage in dozens of behaviors over the course of a match. On a raw data level, what the analysts see are attacks targeting other entities (either other players or non-player characters), items being used, entities being killed, or players dying. But it is the when and where behind these actions that define what they are on a strategic level. If hero A kills hero B in lane, during the early game, surrounded by creeps (a type of non-player character), then it is simply a \textit{one-on-one kill}. If hero A kills hero B by entering his lane from the side jungle, and hero A has an ally in lane who drew hero B's attention to set them up for the attack, then it is a strategic behavior known as a \textit{Gank}. Hero A can also kill hero B in the midst of a team fight, because hero B is the enemy team's primary damage dealer and hero A wants to remove them from play. In all three cases, hero A killed hero B, and without the contextual details, it is unlikely that these kills will be classified correctly. This may have implications on how the system recommends teams, items or guides the player through a game. 

\section{Addressing Ethical Concerns}
Data is a powerful tool for game development that allows designers to understand players and increase engagement and profit at a massive scale. As discussed above, there are several ethical issues discussed concerning the use of data, these can be summarized as follows:
\begin{enumerate}
    \item \textit{Retention strategies that cause addiction:} many of the data driven techniques meant to keep players invested can lead to uncontrolled, addictive gameplay habits. 
    \item \textit{Monetization techniques that encourage irresponsible spending:} placing game content and bonuses behind paywalls often motivates players to spend money they may not have to continue progressing.
    \item \textit{Lack of data on specific groups:} there may be a lack of data on specific groups, such as non-expert players who may not produce enough data throughout gameplay to allow us to address this group or model their behaviors. 
    \item \textit{Lack of models that capture individual differences:} even with some data, if we model based on all players' behaviors or with the big group rather than splitting the data into smaller subsets, we may marginalize specific groups of people. 
    \item \textit{Lack of data capturing details of context:} most game data are often aggregated statistics that may not capture decisions made moment to moment or the context behind such decisions. 
    \item \textit{Lack of transparency or interpretability of the models or results:} we often produce models using black-box methods that do not allow us to interpret the reasons behind the results produced. 
\end{enumerate}

These issues are relevant to data in general, and in the areas of machine learning and data science at large, there has been some work addressing problems 3, 4, and 6 above. Work addressing 3 has looked into selecting data subsets that are better representative of the sample to minimize bias \cite{mckenzie2015subset}. Work addressing 4 has examined using methods that are adjusted and tuned to better capture differences between individuals \cite{huang2006combining,navarro2006modeling,recker1995modeling,viken2002modeling}. The field of Explainable AI seeks to address issue 6 through detailed study of how AI systems can be explained to humans \cite{doran2017does,miller2017explainable,wang2019designing}. However, some of these issues, especially 4 and 5, are of greater pressing concern to games due to their open, dynamic environments. Issue 6 is also of particular concern if one wishes to increase interpretability to ensure player data is understood correctly.

The first three issues, addiction, monetization, and marginalization, are, for the most part, not issues that can be addressed through novel methods for data analysis. The first two deal more with the ethical application of data, rather than its analysis, while the third relies on the ethical collection of data. In this section, we will focus specifically on the latter three issues: individual differences, context, and interpretability, which we believe can be addressed through methodological approaches.

\subsection{Addressing Individual Differences}

%groupings (Simpson's Paradox) and
To address the issue of individual differences discussed above, Canossa et al. \cite{canossa2018like}, in their work on \emph{The Division} \cite{division,canossa2018like}, used a multi-step approach involving expert designers interpreting and grouping players. They first clustered player data and then presented the resulting clusters to the game designers allowing them to assign meaning to the clusters. If the clustering results were not meaningful, different clustering techniques were used. This enforced the idea of data used to guide and inform, but not make decisions. It also allowed designers to be the judge of figuring out the different groupings in their data rather than treating the data of all the players as one whole. Their second step was to use sequence analysis to analyze each clusters' sequence of actions \cite{canossa2018like}. While this is a good step towards addressing differences, the focus on clustering means there is still a risk of losing individuals who sit too far from the clusters' centers.

Another approach used and proposed by several researchers is the use of visualization to seek meaning in players' behaviors. In addition to its benefits to preserving context and providing interpretable models, it also helps with the preservation and analysis of individual differences. Several visualization systems have been proposed \cite{gagne2011deeper, moura2011visualizing, wallner2012spatiotemporal}. Most of these visualization systems used a game map to show how players moved and exhibited actions through their play. Visualization proved beneficial in preserving individual differences in game data \cite{wallner2013visualization}. 

Visualizing data makes it possible to easily see not only players who are similar to each other (who form the clusters), but those who are dramatically different from each other, and those who don't quite fit into any particular cluster. However, as more player data is displayed such individual traces become cluttered making it hard to decipher meaning. To remedy this, several researchers explored the visualization of clusters based on the sequences displayed. A good example of this is \emph{Glyph} \cite{nguyen2015glyph}, where researchers showed players' actions using two synchronized windows: one showing the sequences of players' action and the other showing such sequences clustered by similarity. Using such a visualization was effective in allowing designers to see individual differences in how players solved problems or behaved through time, because the clusters clearly showed popular strategies or patterns and uncommon ones.  

Additionally, Ahmad et al. \cite{ahmad2019modeling} presented a mixed methods approach in which action sequences, a human-in-the-loop, and two visualization systems are used to preserve spatio-temporal context, and facilitate the analysis of individual differences. They use one visualization system, \emph{Stratmapper}, to present the actions of players in the context of when and where in-game they occurred, preserving a greater amount of situational factors than traditional techniques. This allows humans to create behavioral labels based on domain knowledge, which allows the abstraction applied to the data to reflect how players think about gameplay \cite{ahmad2019modeling}. The labeled data is then exported to a second visualization, \emph{Glyph}, where individual action sequences and clusters of sequences can be seen simultaneously. This allows for the analysis of individual differences, as individual variations can be easily identified within the clusters, and their action sequences (and how they differed) can be viewed \cite{ahmad2019modeling}. 

\subsection{Addressing Context}

While Canossa et al. \cite{canossa2018like} didn't address problem 5 (situational context) completely, their use of sequences preserves some aspect of the action and some context of what happened before the action was taken. They showed that this technique was superior to using aggregate data \cite{drachen2012guns}. To address the further issue of context (problem 5), Aung et al. incorporated both temporal and spatial context while analyzing player telemetry data from \emph{Just Cause 2} \cite{jc2,aung2019trails}. They then used clustering approaches and allowed experts to establish meaning for the clusters by examining the trajectories of the players closest to the centeroid of each cluster \cite{aung2019trails}. While this technique is effective at incorporating situational context into behavioral profiling, examining just the centroid of the clusters marginalizes individual differences.

Additionally, much of the works that leverage visualization towards observing individual differences also successfully preserved a greater deal of contextual information through the ability to observe data in the spatial context of the game map \cite{ahmad2019modeling,wallner2013visualization,wallner2012spatiotemporal,gagne2011deeper,moura2011visualizing}. As an illustrative example, Ahmad et al.'s \cite{ahmad2019modeling} work, discussed above, provides a spatio-temporal visualization that allows human analysts, and domain experts, to examine the progression of in-game events in the context of where they occurred on the map. The experts can then create data labels that reflect this spatio-temporal context, ensuring that the the data visualized in \textit{Glyph} will include key contextual components. This increases the chances that the sequences will be interpreted correctly by researchers \cite{ahmad2019modeling}.

\subsection{Addressing Transparency and Interpretability}

Beyond the scope of games, eXplainable AI (XAI) is a field dedicated to highlighting the dangers of opaques systems, and developing algorithms that are interpretable and explainable \cite{doran2017does,holzinger2018machine}. They emphasize that there is value in ensuring interpretability on the user end as well as among the engineers \cite{miller2017explainable}. There is not much work applying XAI to game data, thus it is still an open area of exploration to see how XAI practices could be applied to game data analytics to improve transparency and interpretability. However, Zhu et al. \cite{zhu2018explainable} present a detailed vision of how XAI could enhance AI related systems in the domain of games. Specifically, they give examples of how XAI can be applied to AI assisted co-creation systems, procedural content generation systems, and systems that can evaluate in-game AI agents \cite{zhu2018explainable}. 

Game data is complex and difficult to interpret, and while it is not practical to collect think aloud data in the wild, being able to expose models to players, and allowing them to comment on or correct those models, can increase their transparency (addressing problem 6 above). Such an approach can also increase their accuracy, as discussed by Mikkelsen et al. \cite{mikkelsen2017ethical}. 

Work has shown the benefits to including the players' thoughts in the modeling process. Horn et al. \cite{horn2016opening} used a mixed-method approach to analyzing player's behavior and strategies in an educational game. They used play traces as well as think-aloud utterances in their analysis. Specifically, they analyzed player's progression by hand as well as used hierarchical clustering to cluster play traces. This resulted in both questions and conclusions regarding player's strategic thinking that they were then able to answer or confirm by analyzing player session think-aloud data, and connecting what players said with the moves they made \cite{horn2016opening}. While this type of analysis is time consuming and qualitative, it does allow the analyst to be involved in the modeling process and addresses issues of interpretability. 

The work of Nguyen et al. \cite{nguyen2015glyph} and Ahmad et al. \cite{ahmad2019modeling} discussed above also address transparency. Through their visualization systems, they facilitate human-in-the-loop modeling by facilitating human interpretation through visualization and relying on human reasoning to draw conclusions from the visualized data \cite{nguyen2015glyph,ahmad2019modeling}. A number of other works also draw on visualization techniques to increase the human interpretability of data \cite{wallner2013visualization,wallner2012spatiotemporal,wallner2019aggregated}.

All of the approaches discussed above have merit, and all successfully address at least some of the problems discussed above. However, none are able to address all concerns, and some concerns have yet to be sufficiently addressed. Further, while the techniques above discuss some early directions, we are still far from finding practical approaches that scale and address these ethical concerns. Therefore, further exploration is still needed in this area.

\section{Takeaways}

As we have illustrated in the previous section, existing work has explored new and innovative methodological approaches to data driven game design. Many of these approaches help address a number of the ethical concerns that we have outlined. In this section, we will summarize the contributions of previous work and provide takeaways that developers and researchers can use to guide their procedures. Further, we will summarize open problems as they relate to each ethical concern. A synopsis of these topics can be seen in Table \ref{tab:summaryTable}.

\begin{table*}[]
    \centering
    \begin{center}
    \begin{tabular}{ |p{4.5cm}|p{7cm}|p{5cm}| } 
     \hline
     \textbf{Ethical concern} & \textbf{How it's been addressed} & \textbf{Open problems} \\ 
     \hline
     \hline
     Retention and Addiction & 
     Previous work presents conflicting evidence regarding the extent to which video games can become addictive. However, as a precaution, it is suggested that designers evaluate how they use data to inform design and encourage players to continue playing with a critical eye. & 
     More work is needed to determine whether or not video game retention techniques cause addictive habits, vs. external factors beyond the design of the game. Additional work is also needed to determine the best way to encourage gameplay while preventing damaging behavior.
     \\
     \hline
     Monetization &
     Previous work recommends being transparent with players regarding the presence of paid content or monetization models in games, such that players can choose to avoid such content if exposure would be financially unsafe. & 
     Such an approach, while honest, is at odds with the goal of making a profit, and further work is needed to determine what regulation and presentation practices can best balance these two factors. 
     \\
     \hline
     Lack of Data & 
     Previous work recommends regulation or policy changes to better guide how data is collected and organized making sure that enough data is collected from under represented minorities when possible. &
     There is no clear way to ensure that enough data is collected from all groups, especially in the case of gameplay data, where novice players may play significantly less than experts, and may not produce enough data at all.
     \\ 
     \hline
     Individual Differences & 
     Previous work showed that individual differences in games is not just a demographic difference, but can be about different style of play, or connected to game expertise and personality. To address and capture individual differences, three methods are recommended: (1) Developing models on the different player behavioral clusters capturing variations in players' behaviors, (2) using sequential rather than aggregate data can capture granular individual differences, and (3) visualizations can be used to represent the data where individual differences are visible or highlighted. &
     While these methods have been more effective at capturing individual differences than traditional techniques, more studies are needed to examine the exact extent to which each technique is able to address concerns of marginalization.
     \\
     \hline
     Context & 
     Previous work have shown importance of context in the analysis process. It is then recommended that analysts identify what contextual elements are important to the data and leverage data formats and presentation tools to encode that context into the data representation. &
     It is difficult to define what constitutes context, and how it can best be derived from game data. Including context also increases the complexity of already incredibly complex data.
     \\
     \hline
     Transparency and Interpretability & 
     Previous work have shown the importance of transparency and interpretability and discussed three ways to accomplish this: (1) using mixed methods leveraging qualitative techniques to embed labels or knowledge into the modeling process, (2) using visualization to allow for better representation and transparency, and (3) using explainable AI to explain resulting models from black-box machine learning techniques. &
     More work is needed to determine the best practices for visualization and scalable mixed methods within the domain of game data. Additionally, there is space for work exploring the applicability of explainable AI techniques to the game data domain.
     \\
     \hline
    \end{tabular}
    \end{center}
    \caption{The six identified ethical issues with game data analytics, how they are addressed in the literature, and open problems.}
    \label{tab:summaryTable}
\end{table*}

Previous work has shown that there are concerns regarding retention and monetization. Many games leverage player data to increase playtime or spending, and there is concern that this practice can cause addiction or financially irresponsible decisions \cite{mikkelsen2017ethical,hodent2019cognitive}. Previous work has suggested that, in both cases, designers think critically about the potential harm that could be caused by using data towards such a goal. Further, with regards to monetization, it is encouraged that developers be upfront about paid content, and models that track spending, such that players can avoid games that may entice them into bad practices. However, with regards to both addiction and monetization concerns, these suggestions are somewhat at odds with the goal of turning a profit. Thus, more work is needed to determine the best practices to encourage safe gameplay without harming a game's success.

Previous work have shown that there is a lack of data pertaining to certain groups, which, as discussed above causes marginalization \cite{o2016weapons}. This is not necessarily a concern that can be addressed through new techniques. As such, little work has addressed this through methodological approaches. Instead, we believe that this could be addressed through regulations that govern the collection and organization of data. Although there is little exploration of the viability of solutions, previous works cite the concern, and could be used as a reference to guide policy \cite{mikkelsen2017ethical,bostrom2014ethics}. However, this is still an open problem especially since it is difficult to collect more data from non-expert players who don't usually have enough gameplay data due to being novice or non-expert. Further, collecting more data to capture behavioral variations in style or personality is hard.

Related to lack of data, previous methods face challenges accounting for individual differences as discussed above. This is one of the most pressing concerns for games, as game data is vast and varied. Existing work has demonstrated the benefit of doing analysis on clusters of players separately rather than the whole population. Further, previous work have also shown that using sequential data \cite{canossa2018like,nguyen2015glyph} over aggregates \cite{drachen2012guns} can prevent the marginalization of behavioral variations that may not be common. Additionally, existing work leveraged visualization \cite{nguyen2015glyph,ahmad2019modeling,gagne2011deeper, moura2011visualizing, wallner2012spatiotemporal} as a way of making all of the players' trajectories, no matter how clustered or disparate, visible to the analyst. While these methods have been shown to capture individual differences, we don't have scientific evidence or a study that discusses marginalization and behaviors variations and methods the best capture them. Thus, more work is still needed perhaps expanding on sequential data analysis, cluster analysis and visualizations with more studies targeting the issue and question of marginalization and individual differences. 

Context-aware analytics, as discussed above, is another important issue especially for game data science. This is due to the importance of environmental factors and their impact on players' behavior. Again, sequential data has demonstrated its ability to capture temporal context \cite{canossa2018like} and can be combined with spatial data to also capture location based context \cite{aung2019trails}. Here too, visualizations are being leveraged to allow analysts to easily capture the spatio-temporal context of the data \cite{ahmad2019modeling}. The primary takeaway, similar to the previous issue, is to consider what elements of the game environment are of primary concern to the data being examined (such as time and location) and leverage data formats (such as sequences) and visualizations (such as heat maps) as ways to encode such context into the data representation. However, there are still open problems to this issue, including what constitutes a virtual context? How can it be derived from game data? Additionally, including context means increasing an already high-dimensional data, and thus this adds the issue of dealing with high dimensional analysis space.  

The lack of interpretability or transparency of modeling techniques such as black-box machine learning techniques is an important issue echoed above. Existing work in games demonstrated the benefit of mixed methods approaches in linking players' reasoning to their actions, ensuring that their data is interpreted correctly by researchers \cite{horn2016opening}. Similarly, visualizations have proven useful in increasing the transparency of data, such that it can be more easily interpreted and more accurately understood \cite{ahmad2019modeling,nguyen2015glyph,wallner2019aggregated}. Beyond the domain of game data, explainable AI has dedicated much time and resources towards achieving the goals of transparency and interpretability \cite{doran2017does,miller2017explainable}, and existing work has shown potential in applying XAI concepts to games \cite{zhu2018explainable}. However, it has yet to be seen how XAI can guide more ethical game data science. 

Going forward, we encourage work in game data analytics to explore such mixed methods and explainable AI approaches to ensure more interpretable and transparent data. Additionally, we encourage the exploration of exposing the model to players, which could potentially increasing player trust in the collection of their data \cite{mikkelsen2017ethical}. The open problems in this area is centered on the need for more work around the development of methods for visualization, mixed-methods approaches that scale as well as explainable AI techniques. 

\section{Conclusion}

Game data science and Artificial Intelligence algorithms are powerful tools that can be used effectively within software development. However, they also have the capability of inflicting damage, when implemented improperly, trained on incomplete or biased data, or when decisions they make are obscured or hidden \cite{o2016weapons,bostrom2014ethics}. Game research and production processes, where such data-driven work has bloomed in the last few years \cite{el2016game}, are not immune to these dangers. Game data is incredibly complicated. In addition to the information it contains, it is also informed by various external contexts that impact the way players engage with gameplay. In this paper we identified several data analysis problems, including lack of data, lack of models that capture individual differences and context, and lack of transparency as underlying issues causing several ethical concerns and potential dangers to players such as predatory monetization, marginalization, and misrepresentation \cite{mikkelsen2017ethical}. 

%These ethical concerns do pertain to AI, ML, and data driven decision making at large, however, the highly complex nature of game data further complicates several of them. Specifically, while the issues of lack of data and transparency are more general, the inability to capture individual differences and context are especially important to game data. Game data, representing player actions taken in highly dynamic environments with numerous external influences and multiple possible progression trajectories, can be easily misinterpreted if removed from its context. Additionally, because there are so many ways for a player to progress through a game, there are likely to be more outliers within the data. If these outliers are overlooked during analysis, it can lead to misinformed, or even harmful, game design updates. This can cause incorrect changes in adaptive systems, player frustration, and decreased player retention. 

We further outlined several promising directions to address these issues. It should be noted that this area has received much attention in AI and ML lately. However, much of the work within AI and ML either focused on algorithm design rather than other aspects of human engagement, didn't address the issues we discussed here with game data, or were not discussed or applied to the game data science. In game data science research, there are several approaches that showed great promise. These involved increasing human involvement in all data science and analysis stages, increasing the interpretability of the results, and the use of visualization to preserve context, capture individual differences, and increase transparency of models. However, such approaches are still exploratory and face challenges in terms of scale and practicality. More work is needed to engage the community, as well as ethicists and social scientists, in discussing various new avenues for methodologically tackling the ethical problems and their root causes discussed in this paper. Further, there are probably other causes and problems that we can discuss or uncover as we start to tap this important area of research. 

%%%%%%% PAPER ENDS HERE %%%%%%%

\bibliographystyle{ACM-Reference-Format}
\bibliography{sample-base}

\end{document}